\documentclass[12pt]{article}
\usepackage{latexsym}
\usepackage{epsfig}
\usepackage{amsfonts}
\begin{document}

 \title{\bf On High Energy Scattering in Extra Dimensions}
\author{D.I.Kazakov$^{1,2}$ and G.S.Vartanov$^{1,3}$}
\date{}
\maketitle \vspace{-0.8cm}
\begin{center}
{\it $^1$Bogoliubov Laboratory of Theoretical Physics, Joint
Institute for Nuclear Research, Dubna, Russia \\[0.1cm]
$^2$Institute for Theoretical and Experimental Physics, Moscow,
Russia\\[0.1cm]$^{3}$Moscow Institute of Physics and Technology, Moscow, Russia}
\end{center}

\begin{abstract}
We analyze the behaviour of the high-energy scattering amplitude
within the brane world scenario in extra dimensions. We argue that
contrary to the popular opinion based on the Kaluza-Klein
approach, the cross-section does not increase with energy, but
changes the slope close to the compactification scale and then
decreases like in the 4-dimensional theory. A particular example
of the quark-antiquark scattering due to the gluon exchange in the
bulk is considered.
\end{abstract}

\section{Introduction}

Extra dimensional theories~\cite{loc} have attracted considerable
attention in recent years.  Various brane world models provide
wide possibilities for phenomenological applications (For review
see, e.g.~\cite{review}). However, the lack of A consistent field
theory in extra dimensions compels one to stick to the
Kaluza-Klein approach at the tree level and assume that the string
theory cures the problems with divergences at high energy.

One of the immediate consequences of the K-K approach is the
increase in the scattering cross section with energy due to the
exchange of an infinite tower of K-K modes~\cite{xsec}. To get a
finite result, one usually needs to introduce some cutoff, thus
making the amplitude essentially cutoff dependent. The cutoff may
well have a physical meaning, for instance, representing the width
of the brane,  but this does not reduce the amount  of
arbitrariness. Assuming that in the full theory this dependence
disappears one can try to look for a qualitative picture.

Yet the increase in the total cross section being attractive from
the point of view of observations cannot last forever due to the
breaking of unitarity, so at some energy scale this behaviour has
to be changed.

We try to analyze this situation following the conception of the
low-energy effective theory  based on the Wilson approach to the
renormalization group~\cite{Wilson}. Though it does not allow one
to get a fully consistent theory, it has an advantage of
possession of the effective low-energy theory without cutoff and
has some remarkable properties at the fixed point~\cite{K1,K2}. We
argue that taking into account the radiative corrections in extra
dimensions the behaviour of the cross section is changed and
finally leads to decreasing dependence on energy. However, this
theory is also valid only up to some energy scale.

To illustrate the main features  of an effective theory and to
compare it with the K-K approach, we consider below the
quark-antiquark scattering due to gluon exchange in the bulk.

\section{Quark-antiquark scattering in the brane world}

\begin{figure}[htb]
\begin{center}
  \leavevmode
  \epsfxsize=2.5cm
 \epsffile{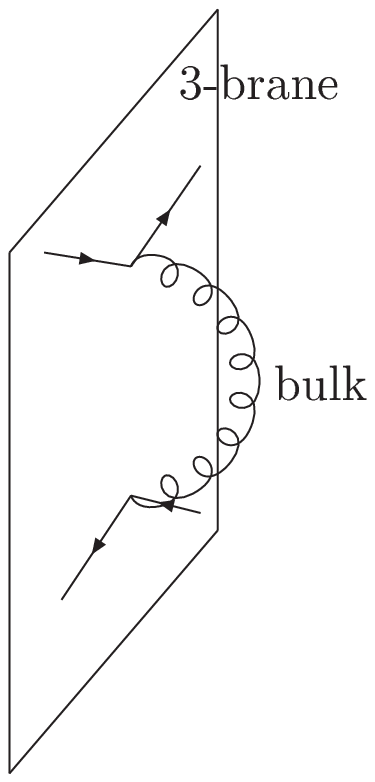}
\end{center}\vspace*{-0.5cm}
\caption{Quark-antiquark scattering due to gluon exchange in the
bulk} \label{brane}
\end{figure}

Assume now that the quark fields are localized on the
3-dimensional brane of zero width  while gluons can propagate in
the bulk of $d$ extra dimensions. Schematically, the situation is
shown in Fig.\ref{brane}.

\noindent The interaction term on the brane looks like~\cite{loc}
$$g_{4+d}\int d^4xd^dy \bar \psi(x)T^a\gamma^\mu A_\mu^a(x,y)\psi(x)
\delta^d(y),$$ where the coupling $g_{4+d}\sim 1/M_c^{d/2}$, $M_c$
being some characteristic scale (scale of compactification of
extra dimensions, localization scale, etc.).

 At the tree level the corresponding scattering
amplitude in the $s$-channel can be written as
\begin{equation}\label{amp}
  Amp\ \sim\   \bar u(p_1)\gamma^\mu T^au(p_2)\ g_{4+d}^2
  \frac{(-)^{d/2}}{(2\pi)^d}\int d^dp
  \frac{1}{s-p^2+i\varepsilon}\
  \bar u(p_3)\gamma^\mu T^au(p_4),
\end{equation}
 The Euclidean integral over additional momenta  in eq.(\ref{amp}) may be
discrete if  extra dimensions are compactified. Then one has an
infinite sum over the Kaluza-Klein modes with increasing mass.

If $d>1$, then the integral (or a sum) in eq.(\ref{amp}) is
divergent. Usually, one introduces an ultraviolet cutoff $\Lambda$
which may be understood as inclusion of the nonzero width of the
brane or some other way. This leads to arbitrariness of
predictions and depends on unknown details of localization.

With the cutoff the integral takes the form~\cite{xsec}
\begin{eqnarray}\label{sum}
 I_d &=& g^2_{4+d}\Omega_{d}\frac{(-)^{d/2}}{(2\pi)^d}
\int^{\Lambda}_0 \frac{p^{d-1}dp}{s-p^{2}+i\varepsilon}
\\ & & \hspace*{-1.3cm} =\frac{g^2}{M_c^{2}}
\frac{1}{(4\pi)^{d/2}\Gamma(d/2)} \left\{
[i\pi+\log(\frac{s}{\mu^2})]\left( \frac{s}{M_c^{2}}
\right)^{d/2-1}\hspace{-0.3 cm} + \sum_{k=1}^{[(d-1)/2]} c_{k}
\left( \frac{s}{\Lambda^{2}} \right)^{k-1} \left(
\frac{\Lambda^2}{M_c^2} \right)^{d/2-1} \right\},\nonumber
\end{eqnarray}
where we have substituted $g^2_{4+d}=g^2/M_c^d$ and $
\Omega_{d}=2\pi^{d/2}/\Gamma(d/2)$. Equation(\ref{sum}) replaces
the usual $g^2/s$ behaviour in 4 dimensions.

In the leading order one has
$$I_d=\frac{g^2}{(4\pi)^{d/2}\Gamma(d/2)}\frac{1}{M_c^2}\left[i\pi+\log(\frac{s}{\mu^2})\right]
\left( \frac{s}{M_c^{2}} \right)^{d/2-1}. $$

To glue it to the low energy 4-dimensional theory  at $s=M_c^2$,
we adjust the scale $\mu$ and finally get
\begin{equation}\label{KKprop}
I_d=\frac{g^2}{(4\pi)^{d/2}\Gamma(d/2)}\frac{1}{M_c^2}\left[\log(\frac{s}{M_c^2})
+(4\pi)^{d/2}\Gamma(d/2)\right] \left( \frac{s}{M_c^{2}}
\right)^{d/2-1}.
\end{equation}

It is easy to see that expression (\ref{KKprop})  leads to
increasing cross-section. Indeed, the differential cross-section,
which in the 4-dimensional case looks like
\begin{equation}
    \frac{d\sigma^{(4)}}{dt}=
\frac{g^4}{4\pi}\frac{1}{9s^2}\left\{
\frac{t^2+u^2}{s^2}+\frac{u^2+s^2}{t^2}-\frac 23
\frac{u^2}{ts}\right\} \label{x4},
\end{equation}
now  takes the form (for $s,|t| \geq M_c^2$)
\begin{eqnarray}
 \frac{d\sigma^{(4+d)}_{KK}}{dt}&=&
\frac{g^4}{4\pi}\frac{1}{9s^2}\frac{1}{(4\pi)^{d/2}\Gamma(d/2)}
\left\{ \frac{t^2+u^2}{M_c^4}\left(\frac{s}{M_c^2}\right)^{d-2}
\log^2(\frac{s}{\widetilde{M}_c^2}) \right. \label{xd}\\ &&
\hspace{-2cm}\left.+
\frac{u^2+s^2}{M_c^4}\left(\frac{|t|}{M_c^2}\right)^{d-2}
\log^2(\frac{|t|}{\widetilde{M}_c^2})
+\frac{2}{3}\frac{u^2}{M_c^4}
\left(\frac{s|t|}{M_c^4}\right)^{d/2-1}\log(\frac{s}{\widetilde{M}_c^2})
\log(\frac{|t|}{\widetilde{M}_c^2})\right\},\nonumber
\end{eqnarray}
where $\widetilde{M}_c^2 \equiv M_c^2\
\exp[-(4\pi)^{d/2}\Gamma(d/2)]$ and we have taken into account the
sign of $t$. In the c.m. frame one has $t=-s/2(1-\cos\theta),
u=-s/2(1+\cos\theta)$ and, respectively
\begin{eqnarray}
   \frac{d\sigma^{(4)} (s,\theta)}{d \cos\theta}&=&
\frac{g^4}{4\pi}\frac{1}{108s}\frac{1}{3}\left\{\frac{
35+8\cos\theta+10\cos^2\theta-8\cos^3\theta+3\cos^4\theta
}{(1-\cos\theta)^2} \right\}
\label{x4c},\\
 \frac{d\sigma^{(4+d)}_{KK} (s,\theta)}{d \cos\theta}&=&
\frac{g^4}{4\pi}\frac{1}{36s}
\frac{1}{(4\pi)^{d/2}\Gamma(d/2)}\left\{
[\frac{(1-\cos\theta)^2}{2}+\frac{(1+\cos\theta)^2}{2}]
\log^2(\frac{s}{\widetilde{M}_c^2}) \right.\nonumber
\\ & +&\left. [\frac{(1+\cos\theta)^2}{2} +2]
[\frac{(1-\cos\theta)}{2}]^{d-2}
\log^2(\frac{s(1-\cos\theta)}{2\widetilde{M}_c^2})\right.
 \label{xdc}  \\
&+& \left. \frac{(1+\cos\theta)^2}{3}
[\frac{(1-\cos\theta)}{2}]^{d/2-1}
\log(\frac{s}{\widetilde{M}_c^2})
\log(\frac{s(1-\cos\theta)}{2\widetilde{M}_c^2})\right\}
\left(\frac{s}{M_c^2}\right)^{d}
.\nonumber
\end{eqnarray}

One can see that the total cross section in the K-K approach
increases with energy and has a different angular dependence
compared to the 4-dimensional case. Note, however, that
eq.(\ref{xdc}) is true only in the energy region $M_c<E<\Lambda$
and has to be replaced further by some decreasing function.

\section{The low-energy effective theory in extra \protect\\ dimensions}

We discuss now the low energy effective theory in $D>4$. We follow
the so-called Wilson renormalization group approach~\cite{Wilson}.

Consider first the usual gauge theory in $D$ dimensions
\begin{equation}\label{l}
  {\cal L}=-\frac 14 Tr F^2_{\mu\nu}, \ \ \
  F_{\mu\nu}=\partial_\mu A_\nu - \partial_\nu A_\mu + g
  [A_\mu,A_\nu].
\end{equation}
The fields and the coupling have the following canonical
dimensions:
$$[A]=\frac{D-2}{2}\ ,\ \  \ \ [F]=\frac D2 \ ,\ \  \ \ [g]=2-\frac D2. $$
Thus, for $D>4$ the coupling has a negative dimension, which
defines some intrinsic scale M, and the theory is
nonrenormalizable in the usual sense. This means that the higher
order operators will inevitably be generated in loop expansion.

Following the Wilson approach one has to write down the
renormalization group equation which in principle involves an
infinite number of operators of type
$$F^2, \ \ g^2(DF)^2,\ \ g^4(DDF)^2,\dots, $$
where the coupling $g\sim 1/M^{d/2}$ is the coupling in the 4+d
dimensional theory. Their influence  on a full theory  is crucial,
since they generate an infinite series of counterterms with
arbitrary coefficients.

Not pretending to resolve this problem at high energy we go now in
the infrared direction, i.e., take $E \ll M$. Then the situation
is somewhat simplified. One finds that all the higher dimensional
operators are irrelevant being suppressed by powers of $g\sim
1/M^{d/2}$ and may be ignored at the tree
level~\cite{Wilson,Weinberg}. The only gauge invariant operator
which is left is the one of the lowest dimension, namely,
$F_{\mu\nu}^2$.\footnote{ In the same way one is left with the
$\phi^4$ operator in case of a scalar theory of critical
phenomena~\cite{Wilson}.}

When going to the loop expansion the situation is more tricky.
Formally, all the UV divergences are absorbed into the
redefinition of the higher order operators which are irrelevant at
low energy. As for the finite contributions of these operators to
the Green functions one expects that they are also suppressed
compared to the leading contribution from the lowest dimensional
operator~\cite{Weinberg}\footnote{We have to admit that we have
not seen any rigorous proof and/or any explicit demonstration of
this fact in the literature.}.

Having all these in mind and leaving a single operator
$F_{\mu\nu}^2$ one can write down the RG equation for the
numerical coupling. Since it is dimensionful  it is useful to take
the dimensionless combination
$$\tilde g \equiv g \mu^{D/2-2} \ \ \ \ \Rightarrow \ \ \ \ [\tilde g] =
0,$$ where $\mu$ is some scale.

In what follows we proceed according to the so-called
$\varepsilon$-expansion procedure and consider a theory in $4+d$
dimensions performing renormalization in the vicinity of the
critical dimension $D=4$. Then the RG equation has a simple form
\begin{equation}\label{rg}
 \mu\frac{d}{d\mu}\tilde g =   \beta(\tilde g, ...)=\tilde g(\frac
 D2-2+\gamma_g(\tilde g)),
\end{equation}
where the dots stand for all those irrelevant operators that we
ignore at low energy.  Here $\gamma_g$ is the usual anomalous
dimension calculated within some renormalization scheme.  In
general $\gamma_g(g)$, and hence $\beta(g)$, may depend on $D$
being finite while $D\to 4$. However, in the MS-scheme this
dependence is absent and $\gamma_g$ can be calculated directly in
the critical dimension $D=4$.

Equation(\ref{rg}) has a nontrivial  fixed point  $\tilde g=g^*$.
It is the so-called Wilson-Fisher fixed point~\cite{Wilson} usual
in scalar theories for $D<4$. However, since in gauge theories,
contrary to the scalar case, the beta function in critical
dimension is negative, the fixed point of this kind exists for
$D>4$~\cite{Peskin,K1}. It is ultraviolet stable, the coupling
approaches it when momentum increases.

At the fixed point the theory possesses no scale: the effective
coupling becomes dimensionless~\cite{K1}. Though the value of the
coupling  is unknown,  the value of the anomalous dimension is
known {\it exactly}: $\gamma_g(g^*)=2-D/2$. It is
 not small but is an integer and  includes all the
radiative corrections.

In the vicinity of the fixed point the Green functions exhibit a
power-like behaviour. In particular the gluon propagator in $D$
dimensions behaves like
\begin{equation}\label{pr}
  G(p)\sim \frac{1}{(p^2)^{1-\gamma_A}}
\end{equation}
In the background gauge, which is appropriate for our case since
the beta function here is completely defined by the propagator,
$\gamma_A=\gamma_g$. The latter can be calculated by perturbation
theory or found from eq.(\ref{rg}) at the fixed point.

At the same time, the low energy effective theory may be used in
the limited energy interval $M_c<E<M$, where  $M_c$ is the
compactification scale and $M>M_c$ is the intrinsic scale of a
higher dimensional theory. For higher energies it should be
replaced by a full theory.

\section{Quark-antiquark scattering in the low-energy \protect\\
effective theory approach}

Consider now the cross-section of the quark-antiquark scattering
in the ET approach. It is given by the same diagrams, but now one
has to use the modified gluon propagator due to eq.(\ref{pr}). At
high energy the behaviour of the gluon propagator is affected by
the fixed point since it is UV attractive. However, here we have a
kind of a controversy: our ET is a low-energy one since we ignore
all the higher order operators and we move in the UV direction
where their influence is essential.

Not being able to go beyond the scale $M$ within the ET approach
we can still solve the RG equations in the vicinity of the fixed
point. If the initial value of the coupling is close enough to its
fixed point value, the influence of the fixed point is essential
even at finite energy.

We proceed in two steps. First we consider the coupling equal to
its fixed point value. The advantage is that the anomalous
dimension of the gluon propagator is known exactly to all orders
of perturbation theory and is equal to $\gamma_A(g^*)=-d/2$. Then
the integral (\ref{amp}) becomes
\begin{equation}\label{fp}
  I_d=g_{4+d}^2\frac{(-)^{d/2}}{(2\pi)^d}\int d^dp
  \frac{(\mu^2)^{d/2}}{(s-p^2+i\varepsilon)^{1+d/2}}\ .
\end{equation}
It is convergent now for any $d$ and equals
\begin{equation}\label{int}
I_d=\frac{g^2_{4+d}(\mu^2)^{d/2}}{(2\pi)^d}\frac{\pi^{d/2}}{\Gamma(1+d/2)}\frac
1s.
\end{equation}
Note that for $d=0$ we have the standard answer $I_0=g^2/s$.

At the fixed point
 $$ g^2_{4+d}(\mu^2)^{d/2} \equiv \tilde g^2 = (g^*)^2,$$
and we have
\begin{equation}\label{fpr}
  I_d=\frac{(g^*)^2}{(4\pi)^{d/2}\Gamma(1+d/2)}\ \frac
1s
\end{equation}

 This means
that we have the same behaviour of the cross section as in 4
dimensions! The only difference is the coefficient that depends on
the number of extra dimensions and the coupling at the fixed
point. The latter is unknown but can be calculated in perturbation
theory.

The result is not surprising since it is given essentially by
dimensional analysis: at the fixed point the theory is scale
invariant and a naive power counting is valid.

Consider now the initial conditions when the coupling  is smaller
but close to the fixed point. We have no all loop information now,
but bearing in mind that we are still in perturbative regime, we
take the one-loop approximation. One has
$$\gamma_A(\tilde g^2)=-b \tilde g^2, \ \ \ \
\beta(\tilde g^2)=d/2\tilde g^2-b\tilde g^4.$$
Then the solutions of the RG equations are
\begin{eqnarray}
  \frac{1}{\tilde g^2}& = &  \frac{1}{\tilde g^2_0}\left(\frac{p^2}{p_0^2}
  \right)^{-d/2}-\frac{2b}{d}
  \left[\left(\frac{p^2}{p_0^2}\right)^{-d/2}-1\right], \\
  G_A(p^2)& = & \frac{1}{p^2} \frac{\displaystyle 1}{\displaystyle
  1+\frac{2b}{d}\tilde g_0^2
  \left[\left(\frac{p^2}{p_0^2}\right)^{d/2}-1\right]}, \label{fg}
\end{eqnarray}
where $p$ is a $(4+d)$-dimensional momentum.  Replacing now the
propagator in eq.(\ref{amp}) by (\ref{fg})
 one gets
\begin{equation}
  I_d =g_{4+d}^2\frac{(-)^{d/2}}{(2\pi)^d} \int d^dp \frac{1}{s-p^2+i\varepsilon}
  \frac{\displaystyle 1}{\displaystyle
  1+\frac{2b}{d}\tilde g_0^2
  \left[\left(\frac{s-p^2}{\mu^2}\right)^{d/2}-1\right]}\ .
\end{equation}
This integral is again convergent for any $d$. To get an explicit
expression we calculate it for $d=2$. The result is
\begin{equation}\label{ex}
I_d=\frac{1}{\mu^2}\frac{\tilde g^2}{4\pi(1-b\tilde g^2)}\
\log(1-\frac{1-1/b\tilde g^2}{s/\mu^2}).
\end{equation}
In the limit  $s\to\infty$ one gets
$$ \frac{1}{4\pi b}\ \frac{1}{s},$$
which coincides with (\ref{fpr}) for $\tilde g^2=1/b$ and $d=2$.

 To link with the low-energy 4-dimensional
theory, we normalize it to $g^2/s$ at $s=M_c^2$. This gives the
equation
$$g^2/M_c^2=\frac{\tilde g^2}{4\pi(1-b\tilde g^2)\mu^2}\
 \log(1-\frac{1-1/b\tilde g^2}{M_c^2/\mu^2})
$$
This is a single equation for two dimensionless variables: $\tilde
g^2$ and $\mu^2/M_c^2$.

Taking, for instance, $\tilde g^2=1/2b, b=7/16\pi^2, g^2=2\pi/5$
(this choice corresponds to QCD with 6 flavours and
$\alpha_s(M_c^2)=0.10$), we get the equation
$$x=\frac{10}{7}\ \log(1+x), \ \ \ x\equiv \mu^2/M_c^2.$$
The solution is $x\approx 0.96$.

The same propagator has to be used in the $t$-channel, but again
one has to take into account the sign change.

To get the cross-section, one has  to replace the propagators in
eq.(\ref{x4}) by their expressions in the effective theory
(\ref{ex}) when the argument exceeds $M_c^2$. The fixed-point
cross section is obtained by taking the asymptotic form of the
propagator (\ref{fpr}).

\begin{figure}[htb]
\begin{center}
  \leavevmode
  \epsfxsize=12cm  \epsffile{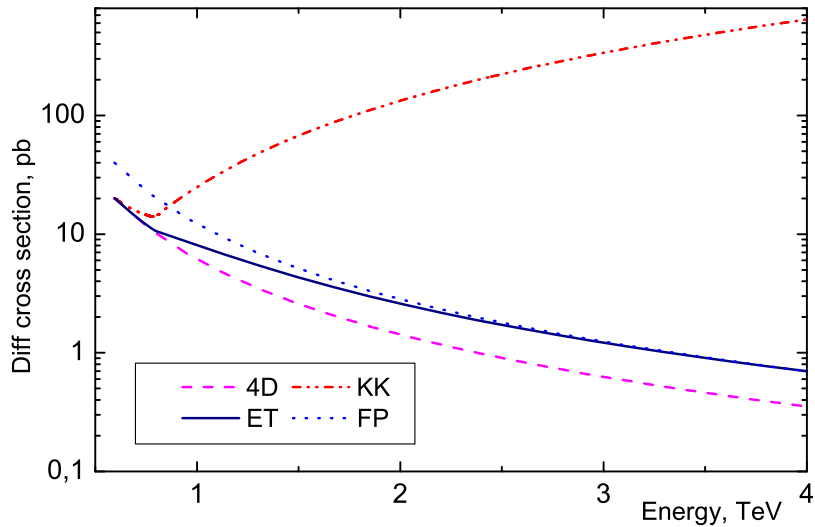}
\end{center}\vspace{-0.5cm}
\caption{The cross section of quark-antiquark scattering in pb for
$D=6$ as a function of energy (solid curve, ET). The dashed curve
(4D) corresponds to the 4-dimensional case, and the dotted one
(FP) is the fixed point limit. The increasing curve is the K-K
prediction (K-K).} \label{cross}
\end{figure}

AS an illustration, we show in Fig.\ref{cross} the resulting cross
section in the case of $d=2$ (or $D=6$) for $\theta=\pi/2$. The
compactification scale $M_c$ is taken to be equal to 1 TeV and the
parameters $b=7/16\pi^2$ and $g^2=0.4\pi$ as above. For comparison
we show also the 4-dimensional cross section (\ref{x4c}) and the
K-K one (\ref{xdc}).

One can see that the cross section in the effective theory
decreases with energy in agreement with unitarity and in
contradiction with the K-K approach. It slowly departs from the
4-dimensional cross section starting from $E=M_c$ by  changing the
slope and finally approaches the fixed-point solution. Note,
however, the range of validity of effective theory.

\section{Conclusion}

In should be emphasized that the problem we discuss has no solid
background due to nonrenormalizability of original higher
dimensional theory. Therefore, any attempts to get the low-energy
results either within the  K-K approach or the ET one inevitably
faces the problem of arbitrariness while handling with the UV
behaviour. The naive cutoff is not a solution to this problem even
if irrelevant operators are suppressed at low energy because the
results for various processes essentially depend on the cutoff
procedure. From this point of view  the effective theory approach
based on a fixed point within the Wilsonian renormalization group
has an advantage of being universal at low energy.

One can see that the cross-section calculated within the ET
approach essentially departs from the one calculated in the
leading order of the K-K approach. The reason is that the
radiative corrections are now power-like and are not the logs as
in 4-dimensional theory. Therefore, their summation drastically
changes the behaviour for increasing momenta. Of course, our
curves have sense only in the IR region; however, as we have
already mentioned, in the vicinity of the fixed point even for
small momenta the behaviour is governed by its attraction.

We considered  quark-antiquark scattering via gluon exchange as an
example to probe the theory at higher dimensions. Apparently the
same situation takes place for other reactions and other exchange
quanta. This means that the increase in the cross-section with
energy which was considered to be the "smoking gun" for extra
dimensional theory, probably is not the case. Based on our
formulas we conclude that the manifestation of possible extra
dimensions is the change of the slope with modified angular
dependence, though the signal is not pronounced and has to be
studied in more detail for a particular process.

\section*{Acknowledgements}
We are grateful to V.Rubakov, A.Slavnov, Y.Okada, and O.Teryaev
for useful discussions. Financial support from RFBR grant \#
02-02-16889 and grant of the Ministry of Science and Technology
Policy of the Russian Federation \# 2339.2003.2 is kindly
acknowledged. D.K.I would like to thank the Theory Group of KEK
where this paper was finished for support and hospitality.

\end{document}